# An Online Data-Driven Method for Microgrid Secondary Voltage and Frequency Control with Ensemble Koopman Modeling

Xun Gong, *Member, IEEE*, Xiaozhe Wang, *Senior Member, IEEE*, and Geza Joos, *Life Fellow, IEEE*

*Abstract*—Low inertia, nonlinearity and a high level of uncertainty (varying topologies and operating conditions) pose challenges to microgrid (MG) systemwide operation. This paper proposes an online adaptive Koopman operator optimal control (AKOOC) method for MG secondary voltage and frequency control. Unlike typical data-driven methods that are data-hungry and lack guaranteed stability, the proposed AKOOC requires no warm-up training yet with guaranteed bounded-input-bounded-output (BIBO) stability and even asymptotical stability under some mild conditions. The proposed AKOOC is developed based on an ensemble Koopman state space modeling with full basis functions that combines both linear and nonlinear bases without the need of event detection or switching. An iterative learning method is also developed to exploit model parameters, ensuring the effectiveness and the adaptiveness of the designed control. Simulation studies in the 4-bus (with detailed inner-loop control) MG system and the 34-bus MG system showed improved modeling accuracy and control, verifying the effectiveness of the proposed method subject to various changes of operating conditions even with time delay, measurement noise, and missing measurements.

## I. Introduction

Microgrids (MGs) are fundamental building blocks of smart grids, whereby distributed energy resources (DERs) are integrated into grids in a scalable fashion. Unlike conventional generators with big rotating masses, power converter-interfaced DERs in MGs lack inertia [1], [2]. In addition, MGs are low- or medium-voltage distribution systems with feeders of low X/R ratios, indicating a coupling between frequency and voltage [2], [3]. Therefore, the frequency and voltage of MGs tend to experience coupled large deviations subject to volatile generation and load, hampering the stable operation of MGs.

For an islanded MG with sufficient capacity, undesired state dynamics can be caused by poorly designed local control, or improper systemwide coordination [2]. There have been many primary control methods (e.g., [4]–[6]) such as virtual impedance damping, modified droop control, and virtual synchronous generator, whereby local stability can be achieved. Nonetheless, the systemwide interaction dynamics emerge as MGs rapidly evolve. MG secondary control is conceived as an effective means to handle the systemwide interaction while it remains challenging owing to the uncertain and nonlinear nature of low-inertia MGs [2], [7].

MG control methods can be categorized as model-based and model-free. The model-based control methods were designed for stability improvement [3], [8], frequency and voltage dynamics control [9], [10], etc. Specifically, Y. Zhang et al [3] assessed systemwide stability and regulated droop gains in a distributed manner based on the dissipative system theory. Likewise, a decentralized control method based on small-signal models was proposed in [8], which improved the systemwide small-signal stability according to the decentralized subsystem designed with Block Gerschgorin Theorem and Lyapunov stability criteria. These works [3], [8] used small-signal models to analyze systemwide dynamics and stability, while a generalized method that can describe large-signal nonlinear dynamics is favorable considering the low-inertia, nonlinear and uncertain nature of MGs. In light of this, A. Bidram et al in [9], [10] proposed multi-agent distributed cooperative control with feedback linearization to deal with nonlinearity, but the method requires accurate dynamic models, parameters, and full state measurement. Generally, all the model-based methods can ensure stability with the best possible performance guarantee based on accurate models and parameters. However, the number of DERs and loads in MGs are increasing and the operating topology and condition can frequently change. Obtaining an accurate model may not always be feasible due to model complexity, high uncertainty (e.g., varying operating conditions and topologies), telemetry errors, etc.

Data-driven model-free control has caught attention to reconcile the concerns of model-based control. Indeed, model-free control methods have been developed in wide-area damping control [11], [12], transient stability control [13], etc. In the context of MGs, H. Zhang et al [14] proposed a data-driven model-free adaptive strategy for the voltage loop of primary control, whereas the estimated system Jacobian for the secondary level is still pre-assumed known. Various modified droop functions targeting primary control were proposed in [3], [5], [6] which improve stability but may not restore system voltage and frequency to their nominal values. Regarding the secondary control, a common model-free method is Proportional-and-Integral (PI) control [7], [15], [16]. However, the conventional PI control may suffer from high starting overshoot, high sensitivity to controller gains and sluggish response to disturbances without proper and frequent control parameter tuning [7]. Another type of MG secondary control

This work was supported by the Fonds de Recherche du Quebec-Nature et technologies under Grant FRQ-NT PR-298827 and FRQ-NT 2023-NOVA-314338.

method widely studied is the averaging/consensus-based secondary droop control [17]–[19] that targets on accurate power sharing in quasi steady-state rather than voltage and frequency stability under large disturbances. To ensure systemwide voltage and frequency stability under both small and big disturbances, a few advanced secondary data-driven control methods have also been proposed recently. Z. Ma et al [7] designed a model-free MG secondary voltage control in which an artificial neural network (ANN) was used to estimate system nonlinear dynamics. Jafari et al [20] and Shen et al [21] also employed ANNs to stabilize the frequency and voltage at the secondary level. Both studies show that machine learning (ML) based methods are effective for performance improvement. However, the universal ANNs and many other learning machines may lack physical interpretability and thus reliability of representing the system's dynamics in diverse topologies and operating conditions. Obtaining abundant offline training datasets that can sufficiently represent the system dynamics could be problematic too. Q-learning, a type of reinforcement learning (RL) technique, was adopted in [22] to help damp system frequency. Nevertheless, RL is a machine learning branch to pursue optimality concerning user-defined rewards, which suffers from the same interpretability and data issues as ANNs.

In this paper, we leverage on Koopman operator theory [23] to design a data-driven MG secondary control method that requires no warm-up training yet possesses theoretical stability guarantee. The Koopman operator theory had been investigated in many previous works [13], [24]–[26]. The authors of [24]–[26] applied the Koopman operator theory for power system nonlinearity modeling, stability assessment, and forced oscillation location, whereas the control inputs were not incorporated. M. Korda [13] investigated the Koopman operator theory with the control inputs for power system transient control, yet the method required offline training data. The theoretical study on Koopman operator control with LQR also gained attention in control community [27], [28]. However, the determination of Koopman embedding functions for the application in power systems of high complexity remains challenging, as it requires a proper pool of lifting function candidates and time-consuming learning that requires abundant offline data. The collection of offline data is another issue for real-world applications especially for new or fast-evolving MGs without knowing the system network information.

To address the challenges of the application of Koopman modeling and control in power systems, we design an online ensemble framework and iterative learning algorithm for the Koopman modeling, which can adaptively weight the estimation from the linear basis and that from the nonlinear basis to provide accurate model estimation in both ambient and transient conditions. As such, the mature optimal linear control method used in power systems---linear quadratic regulator (LQR) can be applied, ensuring well-characterized stability and robustness. The proposed adaptive Koopman operator optimal control (AKOOC) uses only a small window of phasor angle and voltage data from synchrophasors (e.g., micoPMUs) at the DER output ends and can provide effective control to varying MG topologies and operating conditions. To the best of our knowledge, the proposed AKOOC seems to be the first data-driven MG secondary control method that requires no warm-up training yet with guaranteed bounded-input-bounded-output (BIBO) stability and even asymptotic stability under some mild conditions. The advantages of the proposed AKOOC are summarized as follows:

- The proposed AKOOC is purely data-driven using only a small window of synchrophasor data, requiring no knowledge of network topology and parameters and no offline training.
- An ensemble framework and iterative learning algorithm are designed for the proposed AKOOC, which can automatically weight the small-signal and large-signal based models, achieving accurate model estimation and effective control under various operation conditions without the need of event detection or switching between different models (e.g., as in [7]).
- With the proposed AKOOC, the system is guaranteed to be BIBO stable. On top of the BIBO stability, the sufficient condition under which the MG system with the proposed AKOOC is asymptotically stable is also developed.
- The AKOOC is robust to measurement noise, time delay, and missing measurements as tested in numerical studies.

The remainder of the paper is organized as follows: Section II describes the MG system hierarchy and the interface between secondary and primary control. Section III details the proposed control method. Section IV presents case studies for validation. Section V concludes the paper.

## II. MICROGRID SYSTEM DESCRIPTION

The hierarchy of MG control, including the secondary and primary control, is shown in Fig. 1. The secondary controller globally dictates the primary controllers that are responsible for the local stabilization of individual DERs. Each primary controller contains power control, voltage control, and current control from outside to inside loops (see [7] for implementation details). The interfaces of the primary and secondary controllers are at the outer loops of primary controllers, e.g., the power control characterized by conventional droop functions [6]:

Conventional droops: $\begin{bmatrix} \dot{\theta} \\ V - V^* \end{bmatrix} = \begin{bmatrix} \omega - \omega^* \\ V - V^* \end{bmatrix} = \begin{bmatrix} -\sigma_\omega (P - P^*) \\ -\sigma_V (Q - Q^*) \end{bmatrix}$ (1)

with $\quad P_i = \sum_{j=1}^{n} V_i V_j \left( G_{ij} \cos(\theta_i - \theta_j) + B_{ij} \sin(\theta_i - \theta_j) \right)$ (2)

$\quad Q_i = \sum_{j=1}^{n} V_i V_j \left( G_{ij} \cos(\theta_i - \theta_j) - B_{ij} \sin(\theta_i - \theta_j) \right)$ (3)

where $\omega^*$ and $V^*$ are the reference angular frequency and voltage. $\theta$, $\omega$ and $V$ are the desired phasor angle, angular frequency, and voltage for local inner-loop control. The parameters $\sigma_\omega$ and $\sigma_V$ are frequency and voltage droop gains, respectively. $P^*$ and $Q^*$ are the reference active and reactive power injections of individual DERs. $P$ and $Q$ are the low-pass-filtered active and reactive power, and specifically, $P_i$ and $Q_i$ in (2) and (3) represent the active and reactive power injected into the MG from the bus $i$. $G_{ij}$ and $B_{ij}$ represent the equivalent conductance and susceptance between bus $i$ and $j$. If PMU measurement is available at the DER output ends such that angle damping can be added, while the first-order voltage droop controller implemented with integrator, the droop functions can be modified as below [3]:

$$\text{Modified droops:} \begin{bmatrix} \dot{\theta} \\ \dot{V} \end{bmatrix} = \begin{bmatrix} -\frac{1}{\tau_\theta}(\theta - \theta^*) - \frac{\sigma_\theta}{\tau_\theta}(P - P^*) \\ -\frac{1}{\tau_V}(V - V^*) - \frac{\sigma_V}{\tau_V}(Q - Q^*) \end{bmatrix} \quad (4)$$

where $\theta^*$ is the reference phasor angle; $\tau_\theta$ and $\tau_V$ are the time constants for phasor angle and voltage; $\sigma_\theta$ denotes the angle droop gain. $\Delta\theta$ and $\Delta V$ are the phasor angle deviations and voltage deviations.

**Secondary control inputs.** The control signals from the secondary controller are $\Delta P^*$ and $\Delta Q^*$, whereby the DER droop control module will update the reference power at each time by

$$P^* = P + \Delta P^*, \quad Q^* = Q + \Delta Q^* \quad (5)$$

**Discretized small-signal physical model for secondary control.** Typically, the secondary controls for MGs are designed based on the small-signal stability model by perturbing the equations (1) and (4) around the equilibrium point [3], from which we can obtain the small-signal physical models (6) and (7) respectively

$$\begin{bmatrix} \Delta\dot{\theta} \\ \Delta\dot{V} \end{bmatrix} \approx \begin{bmatrix} -\sigma_\omega(\Delta P - \Delta P^*) \\ -\frac{\sigma_V(\Delta Q - \Delta Q^*)}{T_s} \end{bmatrix} = \begin{bmatrix} -\sigma_\omega & \\ & -\frac{\sigma_V}{T_s} \end{bmatrix} J \begin{bmatrix} \Delta\theta \\ \Delta V \end{bmatrix} + \begin{bmatrix} \sigma_\omega & \\ & \frac{\sigma_V}{T_s} \end{bmatrix} \begin{bmatrix} \Delta P^* \\ \Delta Q^* \end{bmatrix} \quad (6)$$

$$\begin{bmatrix} \Delta\dot{\theta} \\ \Delta\dot{V} \end{bmatrix} = \left( \begin{bmatrix} -\frac{1}{\tau_\theta} & \\ & -\frac{1}{\tau_V} \end{bmatrix} - \begin{bmatrix} \frac{\sigma_\theta}{\tau_\theta} & \\ & \frac{\sigma_V}{\tau_V} \end{bmatrix} J \right) \begin{bmatrix} \Delta\theta \\ \Delta V \end{bmatrix} + \begin{bmatrix} \frac{\sigma_\theta}{\tau_\theta} & \\ & \frac{\sigma_V}{\tau_V} \end{bmatrix} \begin{bmatrix} \Delta P^* \\ \Delta Q^* \end{bmatrix} \quad (7)$$

where $J$ is the power flow Jacobian matrix:

$$J = \begin{bmatrix} \partial P/\partial\theta & \partial P/\partial V \\ \partial Q/\partial\theta & \partial Q/\partial V \end{bmatrix} \quad (8)$$

$\Delta P$ and $\Delta Q$ are the active and reactive power perturbations, respectively; $T_s$ is the time step of secondary control. According to (6) and (7), the discretized small-signal physical model for secondary control can be represented as:

$$x_{k+1} = Ax_k + Bu_k, \quad \text{with } x_k = [\Delta\theta_k, \Delta V_k]^T \quad u_k = [\Delta P^*, \Delta Q^*]^T$$

$$A = \begin{cases} e^{\begin{bmatrix} -\sigma_\omega T_s & \\ & -\sigma_V \end{bmatrix} J}, B = \begin{bmatrix} T_s\sigma_\omega & \\ & \sigma_V \end{bmatrix}, \text{for droops (1)} \\ e^{\left(\begin{bmatrix} -\frac{1}{\tau_\theta} & \\ & -\frac{1}{\tau_V} \end{bmatrix} - \begin{bmatrix} \frac{\sigma_\theta}{\tau_\theta} & \\ & \frac{\sigma_V}{\tau_V} \end{bmatrix} J \right) T_s}, B = T_s \begin{bmatrix} \frac{\sigma_\theta}{\tau_\theta} & \\ & \frac{\sigma_V}{\tau_V} \end{bmatrix}, \text{for droops (4)} \end{cases} \quad (9)$$

where $A$ and $B$ are the discrete-time state transition matrix and control matrix, respectively. $A$ in the physical model (9) depends heavily on $J$ that requires network parameters $G_{ij}$ and $B_{ij}$ as seen from (8) and (2)-(3). In case of unknown network information and large modeling uncertainty, control designed based on (9) may not work effectively. In this work, we leverage on Koopman operator theory to convert the nonlinear dynamical system (1) or (4) into a linear one in a Koopman state space such that we can identify the system model and then design a data-driven secondary control adaptively online without network information.

## III. DATA-DRIVEN MG SECONDARY CONTROL

### A. Koopman Operator Theory

Koopman operator theory [23] is an increasingly prominent theory, which indicates that a nonlinear dynamical system can be transformed into an infinite-dimensional linear system under a Koopman embedding mapping. Unlike small-signal models locally linearized around a fixed point, the Koopman-oriented linear model is valid for global nonlinearity. Although the Koopman operator is typically infinite-dimensional, one can consider finite-dimensional Koopman invariant subspaces where dominant dynamics can be described. Particularly, given a nonlinear dynamical system with external control: $x_{k+1} = F(x_k, u_k)$, where $x \in \mathcal{M}$ and $u \in \mathcal{U}$ with $\mathcal{M}$ and $\mathcal{U}$ being the manifolds of state and control input, we consider the Koopman embedding mapping $\Phi$ from the two manifolds to a new Hilbert space $\Phi: \mathcal{M} \times \mathcal{U} \to \mathcal{H}$ with a set of embedding functions $\Phi(x, u) = [\Phi_1(x, u), \Phi_2(x, u), ..., \Phi_i(x, u)..., \Phi_p(x, u)]^T$. They lie within the span of the eigenfunctions $\psi_j$, i.e., $\Phi(x, u) = \sum_{j=1}^{N_\psi} \psi_j(x, u) v_j$, where $v_j$ are the vector-valued coefficients called Koopman modes and $N_\psi$ is the number of eigenfunctions. The eigenfunctions satisfy $\mathcal{K}\psi_j(x, u) = \lambda_j \psi_j(x, u)$ with $\lambda_j$ being the eigenvalues. Acting on the span of $\psi_j$, the Koopman operator $\mathcal{K}$ advances the embeddings $\Phi(x, u)$ linearly in the Hilbert space $\mathcal{H}$ as [23]:

$$\Phi(x_{k+1}, u_{k+1}) = \mathcal{K}\Phi(x_k, u_k) = \mathcal{K}\sum_{j=1}^{N_\psi}\psi_j(x_k, u_k)v_j = \sum_{j=1}^{N_\psi}(\lambda_j\psi_j(x_k, u_k)v_j) \quad (10)$$

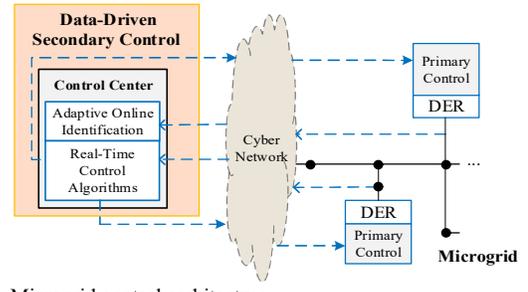

Fig. 1. Microgrid control architecture.

In line with the constant control matrix $B$ of the MG control model presented in (9), we assume: 1) $\Phi_i(x, u) = g_i(x) + L_i(u)$ where $g_i(x)$ is a nonlinear function and $L_i(u)$ is linear with $L_i(0) = 0$ [29]; 2) control inputs are not dynamically evolving to the next step, $\Phi_i(x_{k+1}, 0) = \mathcal{K}\Phi_i(x_k, u_k)$ for all $k$. Then, we have $g_i(x_{k+1}) + L_i(0) = \mathcal{K}g_i(x_k) + \mathcal{K}L_i(u_k) \Rightarrow g_i(x_{k+1}) = \mathcal{K}g_i(x_k) + \mathcal{K}L_i(u_k)$. Note that under the assumption of $\Phi_i(x_{k+1}, 0) = \mathcal{K}\Phi_i(x_k, u_k)$, the Koopman operator is only attempting to propagate the observable functions at the current state $x_k$ and inputs $u_k$ to the future observable functions on the state $x_{k+1}$ but not on future inputs $u_{k+1}$ as inputs are not dynamically evolving (i.e., $[\Delta P^*, \Delta Q^*]^T$ are not state-dependent) [17]. Thus, we will use an equivalent approximation defined in (11) to describe the Koopman-invariant space in a form of extended dynamic mode decomposition with control (EDMDc) [29].

$$\chi_{k+1} = A_C\chi_k + B_C u_k + \delta_{E,k} \quad (11)$$

where $\chi := g(x) = [g_1(x), g_2(x), ..., g_i(x), ... g_p(x)]^T$. $A_C$ and $B_C$ are the state transition matrix and control matrix, satisfying that $A_C\chi_k = \mathcal{K}g_i(x_k)$ and $B_C u_k = \mathcal{K}L_i(u_k)$. $\delta_{E,k}$ is the Koopman modeling error associated with the EDMDc approximation. Different from the linearized model (9) for small signals, the Koopman state space model (11) can also describe large signal-driven nonlinear dynamics. To use this model for control, three consecutive tasks are necessary: (i) finding proper Koopman embeddings (Section B); (ii) establishing a proper Koopman state space model and identifying $A_C$ and $B_C$ (Section C); (iii) applying linear control in Koopman state space (Section D).

### B. Koopman Embeddings

MG systemwide interactions presented in (1)-(4) lead to sinusoidal-driven dynamics that may emerge when subject to large disturbances and low inertia. (i.e., the general solution for the droop-control differential equations contains trigonometric patterns) [30]. Inspired by this, we include the functions $sin\theta$, $cos\theta$, and angular frequency $\omega$ into the Koopman embeddings to describe such underlying dynamics, which were shown effective to describe interaction transients of power grids [13]. Specifically, let $x_k = [\Delta\theta_k, \Delta V_k]^T$, we define the Koopman real-valued embeddings $\chi_k := g(x_k)$ as:

$$\chi_k := g(x_k) = [\Delta\theta_k, \Delta V_k, sin\Delta\theta_k, cos\Delta\theta_k, \Delta\omega_k]^T \quad (12)$$

The augmented embeddings $\chi_k$ advance linearly as presented in (11) with $u_k = [\Delta P^*, \Delta Q^*]^T$. We will need to estimate the matrices $A_C$ and $B_C$ with the measurement data of $x_k$. The estimation is based on the Koopman-based ensemble modeling and learning presented next.

### C. Koopman-based Ensemble Modeling and Learning

Islanded low-inertia MGs may experience both small and large signal dynamics subject to perturbations during operation. This section illustrates a Koopman-based ensemble modeling and learning method to handle the small and large signal-driven dynamics and control adaptively. It includes three parts: (i) An ensemble framework based on the embeddings defined in (12) that combines both small and large signal bases; (ii) A Koopman state space model in the form of (11) derived in the ensemble framework in (i); (iii) An iterative learning method to estimate the model obtained in (ii).

**An Ensemble Framework.** The Koopman embeddings $\chi_k$ can be divided into two subspaces: (i) the subspace $x_k = [\Delta\theta_k, \Delta V_k]^T$ consistent with the structure of the small-signal physical model in (9); (ii) the subspace with supplementary nonlinear basis functions $z_k = \wp(x_k) = [sin\Delta\theta_k, cos\Delta\theta_k, \Delta\omega_k]^T$ to help describe large-signal interaction dynamics characterized by (2) and (3). Similar to the Koopman model predictive control work in [13], [29], an adaptive matrix $C_M$ is used to approximate the mapping $\wp$ between $x$ and $z$:

$$x_k = C_M z_k + e^{(L)} \text{ and } z_k = C_M^\dagger x_k + C_M^\dagger e^{(L)} = \hat{z}_k + e^{(NL)} \quad (13)$$

where $\hat{z}_k = C_M^\dagger x_k$. The terms $e^{(L)}$ and $e^{(NL)}$ denote the modeling errors of the linear approximation between the linear and nonlinear bases $x$ to $z$, satisfying $e^{(NL)} = C_M^\dagger e^{(L)}$. We assume $e^{(L)}$ and $e^{(NL)}$ are bounded. The symbol $\cdot^\dagger$ denotes the Moore-Penrose pseudoinverse of a matrix. $C_M$ is estimated on a rolling basis with the online data of $x$ and $z$. As such, the linear models in (13) can be used to approximate the mapping between $x$ and $z$ with bounded modeling errors.

Consider two Koopman subspace models to be combined: a linear-based model (interpretable for small signals [3]) and a nonlinear-based model (interpretable for large signals [13]):

(linear bases) $\quad \hat{x}_{k+1}^{(\ell)} = Ax_k + Bu_k \quad (14)$

(nonlinear bases) $\quad \hat{x}_{k+1}^{(n)} = C_M z_{k+1} = C_M(A_M z_k + B_M u_k) \quad (15)$

where $A$ and $A_M$ are state transition matrices that are unknown. $B$ and $B_M$ are control matrices for the two sets of bases. $B$ is known as defined in (9) while $B_M$ is an unknown control matrix used to model the impact of the control input $u_k$ on $z_k$. Inspired by the physical equivalence between the Koopman embeddings $x$ and $z$, we assume the control perturbations to the two sets of embeddings keep consistent at any time step $k$, i.e., $Bu_k \approx C_M \cdot (B_M u_k) = (C_M B_M) u_k$. Thus $B \approx C_M B_M$, and we estimate $B_M$ with $B$ by

$$\hat{B}_M = C_M^\dagger B \quad (16)$$

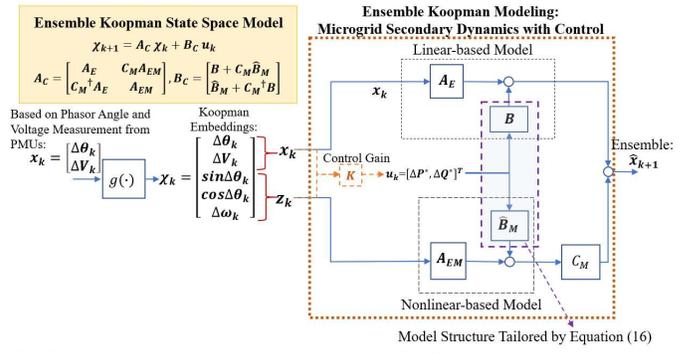

Fig. 2. Koopman-based ensemble modeling of MG secondary dynamics.

Combine the models (14) and (15) with the ensemble weight $\beta$ as below:

$$\hat{x}_{k+1} = (1-\beta)\hat{x}_{k+1}^{(\ell)} + \beta \hat{x}_{k+1}^{(n)}$$
$$= (1-\beta)(Ax_k + Bu_k) + \beta(C_M(A_M z_k + B_M u_k)) \quad (17)$$

It is derived in Appendix A that (17) can be further represented as:

$$\chi_{k+1} = A_C \chi_k + B_C u_k + \delta_{E,k}, \text{ with } \chi_k = [x_k^T, z_k^T]^T,$$

$$A_C = \begin{bmatrix} A_E & C_M A_{EM} \\ C_M^\dagger A_E & A_{EM} \end{bmatrix}, B_C = \begin{bmatrix} B + C_M \hat{B}_M \\ \hat{B}_M + C_M^\dagger B \end{bmatrix} \quad (18)$$

where $A_E := (1-\beta)A - (B - \beta c)K_X$ and $A_{EM} := \beta A_M - (\hat{B}_M - \beta H_M)K_Z$ replacing the original parameters in (17). The definitions of parameters $K_X$, $K_Z$, $H$ and $H_M$ are illustrated in Appendix A. The variable $\delta_{E,k}$ is the Koopman modeling error corresponding to the error term defined in (11). The structure of the reformulated ensemble model is shown in Fig. 2, which can be seen as the ensemble of two models (linear-based and nonlinear-based) that cooperatively describe the MG secondary-level dynamics. Particularly, if we force $C_M = 0$, the large-signal model is opted out and then the ensemble framework is simplified to the small-signal model with the linear basis. The learning of the nonlinear parameters in (17) is now converted to the learning of the new linear matrix parameters $A_E$, $A_{EM}$, $C_M$ and $\hat{B}_M$, which enables fast optimal learning with small data. The details are illustrated next.

**Iterative Learning for Estimating the Koopman State Space Model.** Based on the derived ensemble Koopman model (18), an iterative ensemble learning method is proposed to determine $A_C$ and $B_C$. Consider the dataset $\{x, u\} \in D_k$ collected at time step $k$ in a matrix form:

$$X = \begin{bmatrix} | & | & | \\ x_{k-N}, x_{k-N+1}, \ldots x_{k-1} \\ | & | & | \end{bmatrix}, X' = \begin{bmatrix} | & | & | \\ x_{k-N+1}, \ldots, x_k \\ | & | & | \end{bmatrix} U_k = \begin{bmatrix} | & | \\ u_{k-N}, \ldots u_{k-1} \\ | & | \end{bmatrix} \quad (19)$$

where $x_k$ is the data sample of the state $x$ at time step $k$, and $u_k$ is the data sample of the control input at time step $k$. Calculate the data matrices $Z$ and $Z'$ with $X$ and $X'$ based on the Koopman embedding mapping definition in (12):

TABLE I.
Parameters of Koopman-Based Iterative Ensemble Learning

| **Koopman model parameters to be learned:** |
|---|
| $A_C$ and $B_C$ (i.e., $A_E$, $A_{EM}$, $C_M$, $\hat{B}_M$ to be estimated) |
| **Algorithm parameters:** |
| $B$: control matrix defined in (9). |
| $\eta$: learning rate.        $\varepsilon$: threshold to stop learning. |
| $N_{ITER}$: Maximum iteration number |
| $\gamma$: relaxation constant and $\eta = \gamma/(\gamma+i)$, used to gradually decrease the learning rate for a gradually fine exploitation of local minima). |
| $e$: regression error, defined as: |
| $e = \frac{1}{\sqrt{N \cdot N_r}} \left\| X' - (A_E X + BU) - C_M(A_{EM} X + \hat{B}_M U) \right\|_F$ |
| where, $N$: the sliding window size for regression; |
| $N_r$: the dimension of the state $x$. |

$$Z = \begin{bmatrix} | & | & & | \\ z_{k-N}, z_{k-N+1}, & \dots & z_{k-1} \\ | & | & & | \end{bmatrix}, \quad Z' = \begin{bmatrix} | & | & & | \\ z_{k-N+1}, z_{k-N+2}, & \dots & z_k \\ | & | & & | \end{bmatrix} \quad (20)$$

According to (18), we have

$$X' = A_E X + C_M A_{EM} Z + (B + C_M \hat{B}_M) U + \delta^{(X)} \quad (21)$$
$$Z' = C_M^\dagger A_E X + A_{EM} Z + (\hat{B}_M + C_M^\dagger B) U + \delta^{(Z)} \quad (22)$$

We formulate a least-squares optimization problem to learn the optimal ensemble parameters $W_o = [A_{E,opt}, A_{EM,opt}]^T$ as:

$$W_o = [A_{E,opt}, A_{EM,opt}]^T =$$
$$\arg\min_W \sqrt{\sum_{i_d \in \{k-N,\dots,k-2,k-1\}} \left( \left\| \left( x_{i_d+1} - M_E(x_{i_d}, z_{i_d}, u_{i_d}; W) \right) \right\|_2^2 \right)}$$
$$= \arg\min_W \left\| X' - (A_E X + BU) - C_M(A_{EM} X + \hat{B}_M U) \right\|_F \quad (23)$$

where $\|\cdot\|_2$ represents the 2-norm of a vector and $\|\cdot\|_F$ represents the Frobenius norm of a matrix. Note that state distribution can change substantially after mapping from $x$ to $z$, thus making the ordinary least-squares regression unsuitable for solving the problem (23). To address the issue, we will use iterative learning between $A_E$ and $A_{EM}$ defined in (18). For each iteration, we first assume $A_E$ is fixed, and obtain the analytic solution of the least-squares optimization for $A_{EM}$ by applying the Moore-Penrose pseudoinverse [13] based on (22).

$$A_{EM,opt} = \arg\min_{A_{EM}} \left\| X' - (A_E X + BU) - C_M(A_{EM} X + \hat{B}_M U) \right\|_F$$
$$= (Z' - C_M^\dagger A_E X - (\hat{B}_M + C_M^\dagger B) U) Z^\dagger \quad (24)$$

Then we fix the obtained $A_{EM}$ and calculate the least-squares optimization solution for $A_E$ according to (21)

$$A_{E,opt} = \arg\min_{A_E} \left\| X' - (A_E X + BU) - C_M(A_{EM} X + \hat{B}_M U) \right\|_F$$
$$= (X' - C_M A_{EM} Z - (B + C_M \hat{B}_M) U) X^\dagger \quad (25)$$

It is shown in Appendix B that the ensemble modeling error based on the iterative learning is smaller than the error of individual models [see (51)-(52)]. It should be noted that although the Moore-Penrose pseudoinverse least-squares estimation in (24)-(25) may not provide a statistically optimal solution, the use of least-squares estimation is simple and robust as the information of probability distribution is unknown and Koopman embedding mapping may also change data sample distribution. Hence, the same approach was also adopted in Koopman-based modeling in [13], [24], [28].

The detailed procedure of iterative ensemble learning at time step $k$ based on the least-squares optimization is presented below. Table I summarizes the parameters of the algorithms.

<u>Step 1</u>: Initialize the algorithm parameters in TABLE I. Set $A_E^{(0)} = A_{EM}^{(0)} = I$ at the first time step. Set the iteration number $l=0$.

<u>Step 2</u>: Data preparation at time step $k$:

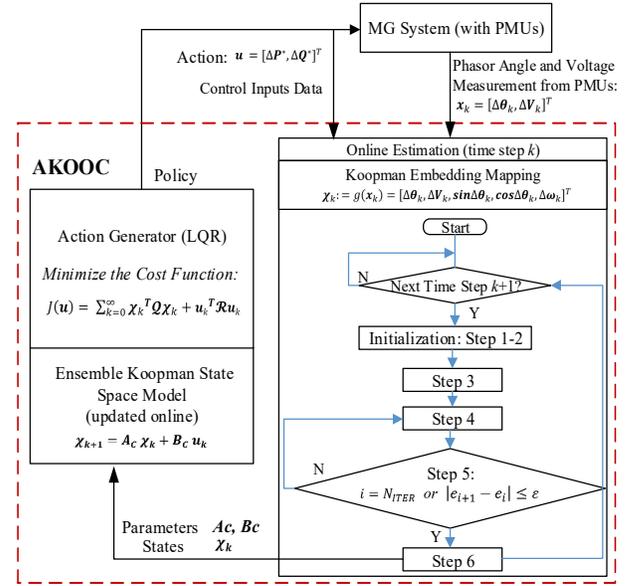

Fig. 3. Overall framework of the proposed AKOOC algorithm.

*With the measured phasor angle, voltage magnitude from PMUs, and the control input dataset from the secondary controller, prepare the data as follows:* $\Delta\theta = [\Delta\theta_{k-N+1}, \Delta\theta_{k-N+2}, \dots \theta_{k-1}]$, $\Delta\theta' = [\Delta\theta_{k-N+2}, \Delta\theta_{k-N+3}, \dots \theta_k]$, $\Delta\theta^- = [\Delta\theta_{k-N}, \Delta\theta_{k-N+1}, \dots \theta_{k-2}]$, $\Delta V = [\Delta V_{k-N+1}, \Delta V_{k-N+2}, \dots \Delta V_{k-1}]$, $\Delta V' = [\Delta V_{k-N+2}, \Delta V_{k-N+3}, \dots \Delta V_k]$ *and* $U = [U_{k-N+1}, U_{k-N+2}, \dots, U_{k-1}]$. *Prepare the data matrices* $X = [\Delta\theta, \Delta V]^T$ *and* $X' = [\Delta\theta', \Delta V']^T$. *Use* $\Delta\theta$ *and* $\Delta V$ *to obtain the data matrices* $Z = \left[ \sin(\Delta\theta), \sin(\Delta V), \frac{\Delta\theta - \Delta\theta^-}{T_s} \right]^T$, $Z' = \left[ \sin(\Delta\theta'), \sin(\Delta V'), \frac{\Delta\theta' - \Delta\theta}{T_s} \right]^T$.

<u>Step 3</u>: *Calculate* $C_M$ *and* $\hat{B}_M$ *by (13) and (16):*
$$C_M = X Z^\dagger, \quad \hat{B}_M = C_M^\dagger B \quad (26)$$

<u>Step 4</u>: *Start the $l^{th}$ iteration. Calculate* $\eta = \gamma/(\gamma+i)$ *and then update* $A_{EM}$ *and* $A_E$ *according to (24) and (25):*
$$A_{EM}^{(l+1)} = (1-\eta) A_{EM}^{(l)} + \eta (Z' - C_M^\dagger A_E^{(l)} X - (\hat{B}_M + C_M^\dagger B) U) Z^\dagger \quad (27)$$
$$A_E^{(l+1)} = (1-\eta) A_E^{(l)} + \eta (X' - C_M A_{EM}^{(l+1)} Z - (B + C_M \hat{B}_M) U) X^\dagger \quad (28)$$

<u>Step 5</u>: *Calculate the regression error:*
$$e_{l+1} = \frac{1}{\sqrt{N \cdot N_r}} \left\| X' - (A_E^{(l+1)} X + BU) - C_M(A_{EM}^{(l+1)} X + \hat{B}_M U) \right\|_F \quad (29)$$

*If $|e_{l+1} - e_l| \leq \varepsilon$ or $l = N_{ITER}$, go to Step 6 and set the initialization values for next time step $k+1$: $A_E^{(0)} = A_E^{(l+1)}$, $A_{EM}^{(0)} = A_{EM}^{(l+1)}$. Otherwise, $l$ increases by 1, and go back to Step 4*

<u>Step 6</u>: *Set* $W_o = [A_E^{(l+1)}, A_{EM}^{(l+1)}]^T$. *Calculate* $A_C$ *and* $B_C$ *as:*
$$A_C = \begin{bmatrix} A_E^{(l+1)} & C_M A_{EM}^{(l+1)} \\ C_M^\dagger A_E^{(l+1)} & A_{EM}^{(l+1)} \end{bmatrix}, \quad B_C = \begin{bmatrix} B + C_M \hat{B}_M \\ \hat{B}_M + C_M^\dagger B \end{bmatrix} \quad (30)$$

*At the next time step, set $k=k+1$. Go back to Step 1 to collect new data and redo the estimation on a rolling basis.*

The iterative ensemble learning can be seen as a cooperative game exploiting the best possible parameters. The theoretical analysis of the error bounds of the iterative ensemble learning is provided in Appendix B. After iteration, the ensemble Koopman state space model (18) is identified with the matrices obtained in (30). Next, we will apply linear optimal control to the adaptively identified model.

### D. Adaptive Koopman Operator Optimal Control

This section uses the identified matrices $A_C$ and $B_C$ in Section III.C to conduct the adaptive Koopman operator optimal control (AKOOC). Fig. 3 shows the overall control framework. We employed the discrete-time linear quadratic regulator (LQR) in

AKOOC to minimize the following objective function at each time step:

$$J(u) = \sum_{k=0}^{\infty} \chi_k^T Q \chi_k + u_k^T \mathcal{R} u_k \quad (31)$$
$$\chi_{k+1} = A_C \chi_k + B_C u_k \quad (32)$$

where $Q$ and $\mathcal{R}$ are cost matrices:

$$Q = \begin{bmatrix} Q_S & \\ & Q_M \end{bmatrix}, Q_S = \begin{bmatrix} q_\theta & \\ & q_V \end{bmatrix}, Q_M = \begin{bmatrix} q_{cos} & & \\ & q_{sin} & \\ & & q_\omega \end{bmatrix}, \mathcal{R} = \begin{bmatrix} r_P & \\ & r_Q \end{bmatrix} \quad (33)$$

where $q_\theta, q_V, q_{cos}, q_{sin}, q_\omega$ are cost submatrices for the states presented in (12). $r_P$ and $r_Q$ are cost submatrices for the control signals $\Delta P^*$ and $\Delta Q^*$. The optimal control input is calculated by:

$$u_k = \begin{cases} U_{LB}, & \text{if } u_k < U_{LB} \\ -K\chi_k, & \text{if } U_{LB} \le u_k \le U_{UB}, \text{ with } K = (B_C^T S B_C + \mathcal{R})^{-1} B_C^T S A_C \\ U_{UB}, & \text{if } u_k > U_{UB} \end{cases} \quad (34)$$

where $K$ is the control gain matrix. $U_{UB}$ and $U_{LB}$ are the upper and lower saturation limits to bound the uncertainty introduced by control inputs. $S$ is the solution of Riccati equation [31]. Given rolling estimation and control, the stability of AKOOC was proved in Appendix C. The sufficient condition under which the system with the proposed Koopman-enabled LQR is asymptotically stable is developed in Appendix D.

## IV. CASE STUDIES

This section presents case studies based on two MG test systems to verify the effectiveness of AKOOC. Although developed based on the droop control functions without local inner-loop converter control, the proposed AKOOC method was tested in a 4-bus MG system in which the converter inner loops were also modeled in detail to demonstrate the effectiveness. The test system was established in MATLAB Simulink 2021b. The DERs in the test system were primary-controlled with the conventional droops in (1). Inner converter control loops (i.e., voltage and current loops) were included and the implementation details can be found in [7]. The interaction of primary and secondary control is therefore conserved in simulation to test the effectiveness of AKOOC in a realistic setup.

After validating AKOOC for the small MG, we did simulations with a larger 34-bus MG test system constructed in PSAT 2.1.10 [32], where the DERs were primary-controlled with the modified droops in (4). To reduce the computation cost for large systems, the converter inner loops are assumed to be able to quickly and accurately track the reference voltage generated by the outer-loop droop control. Therefore, the inner-loop control is not modeled in the simulation model of the 34-bus system.

### A. Control with Measurement Noise, Time Delay, Ambient Perturbation, and Large Disturbance

The small 4-bus MG test system with detailed modeling of inner loops was used first. The network structure was shown in Fig. 4. To mimic real-world MG operation, we added randomized measurement noises, control time delays, and ambient perturbations in the simulation test system. See parameter details in Table II. The secondary control was enabled at 0.8s to restore MG voltage and frequency. Then, a

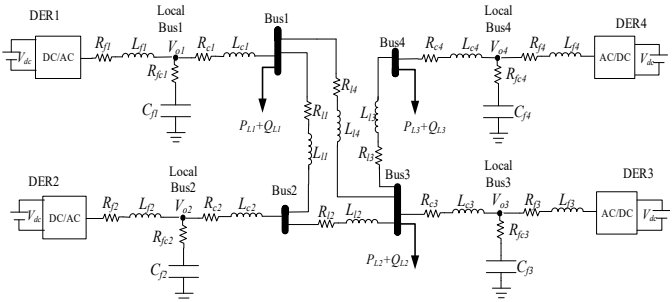

Fig. 4. Microgrid test system I (4 buses, single-line diagram).

TABLE II.
Test System I and Control Parameters

| Parameters | | Values | Parameters | Values |
|---|---|---|---|---|
| Power base | $S_{base}$ | 30kVA | $q_\theta$ | $10^{-5}I$ |
| Voltage base | $V_{base}$ | 480V | $q_V$ | $I$ |
| Primary control time step | $T_{sp}$ | 0.1ms | $q_{cos}$ | 0 |
| Local voltage PI controller | $K_{Pv}$ | 0.5 | $q_{sin}$ | 0 |
| | $K_{Iv}$ | 523 | $q_\omega$ | $10^{-5}I$ |
| Local current PI controller | $K_{Pi}$ | 0.3 | $r_P$ | $I$ |
| | $K_{Ii}$ | 635 | $r_Q$ | $I$ |
| Secondary control time step | $T_s$ | 60ms | $U_{LB}$ | $-\frac{1}{15}$ |
| Constant for learning rate | $\gamma$ | 5 | $U_{UB}$ | $\frac{1}{15}$ |
| Maximum iteration number | $N_{ITER}$ | 150 | Regression window size $N$ | 10 |
| Regression error tolerance | $\varepsilon$ | $10^{-16}$ | | |

Filter and line parameters (in p.u., adapted from [7]):

$R_{f1,2,3,4} = 0.013$, $L_{f1,2,3,4} = 0.663$, $R_{fc,2,3,4} = 0.13$, $C_{f1,2,3,4} = 6.908$, $R_{c1} = 0.010$, $L_{c1,2} = 0.172$, $R_{c3,4} = 0.012$, $L_{c3,4} = 0.221$, $R_{L1} = 0.020$, $L_{L1} = 0.206$, $R_{L2} = 0.046$, $L_{L2} = 0.162$, $R_{L3} = 0.030$, $L_{L3} = 0.270$, $R_{L4} = 0.011$, $L_{L4} = 0.172$.

Load: $P_{L1} = 20kW, Q_{L1} = 9kVar, P_{L2} = 19kW, Q_{L2} = 9kVar, P_{L3} = 12kW, Q_{L3} = 6kVar$

Droop control parameters: DER1-2: $\sigma_\omega = 3.4 \times 10^{-4} Hz/W$, $\sigma_V = 1 \times 10^{-3} V/Var$
DER3-4: $\sigma_\omega = 4.5 \times 10^{-4} Hz/W$, $\sigma_V = 1.5 \times 10^{-3} V/Var$

PMU measurement noise: $\mathcal{N}(0, 0.0056^2)$ (signal-to-noise ratio 45dB [35])

Control time delay: $\mathcal{N}(0.02, 0.002^2)$ s

Ambient perturbation added to the reference of DER output voltage and angle: $\mathcal{N}(0, 0.01^2)$

Note: $\mathcal{N}(a, b)$ is the normal distribution with mean of $a$ and variance of $b$. Control parameters are designed based on Per Unit.

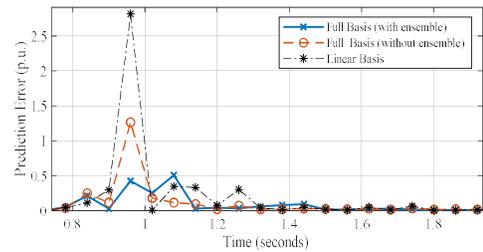

Fig. 5. One-step-ahead prediction error when the ambient perturbation level is $\mathcal{N}(0, 0.01^2)$.

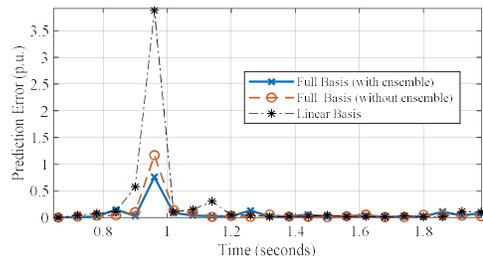

Fig. 6. One-step-ahead prediction error when the ambient perturbation level is $\mathcal{N}(0, 0.02^2)$.

large load disturbance occurred at Bus 3 at 0.9s: $P_{L2}$ increased from 19kW to 79kW and $Q_{L2}$ increased from 9kVar to 39kVar.

***Modeling accuracy of the ensemble Koopman state space model.*** First, we evaluate the modeling accuracy of (12). In line with the regression error defined in (29), let us define the one-step-ahead prediction error for time step $k+1$ by:

$$e_{k+1}^{(pred)} = \frac{1}{\sqrt{\dim[x]}} \|x_{k+1} - \hat{x}_{k+1}\|_2 \quad (35)$$

where $\dim[\cdot]$ represents the dimension of the vector in the bracket. $\hat{x}_{k+1}$ represents the predicted state with the model of interest. For example, the prediction error of the ensemble Koopman state space model is:

$$e_{k+1}^{(pred)} = \frac{1}{\sqrt{2N_{DER}}} \|x_{k+1} - (A_E x_k + BU) - C_M(A_{EM} x_k + \hat{B}_M u_k)\|_2 \quad (36)$$

where $N_{DER}$ represents the number of DERs in the MG. We compared the prediction error of three different ways of modeling: (i) the proposed Koopman-based modeling with the full basis $\chi = [\Delta\theta, \Delta V, \sin\Delta\theta, \cos\Delta\theta, \Delta\omega]^T$ in an ensemble framework; (ii) the Koopman-based modeling with the full basis $\chi$ without the ensemble framework; (iii) the Koopman-based modeling with the linear basis $x = [\Delta\theta, \Delta V]^T$. As shown in Fig. 5, the maximum prediction error of the proposed ensemble Koopman-based modeling using the full basis was smaller than that of the linear basis and that of the full basis without ensemble when subject to the large disturbance at 0.9s. This indicates that the proposed ensemble Koopman-based modeling does not simply combine the linear and nonlinear basis, but also adaptively weights the predictions from the linear and nonlinear bases through the online ensemble framework and the iterative learning. As a result, the proposed ensemble model can provide more accurate prediction. Fig. 6 shows the prediction errors when the ambient perturbation level increased to $\mathcal{N}(0, 0.02^2)$. Likewise, the prediction error with the proposed ensemble Koopman-based modeling using the full basis is smaller than the other models.

***Control results of AKOOC.*** The improved modeling accuracy indicates the potential to improve control. First, the output voltage and frequency trajectories of DERs without applying secondary control are shown in Fig. 7(a). The voltage at Buses 1-4 dropped and the frequency became unstable due to

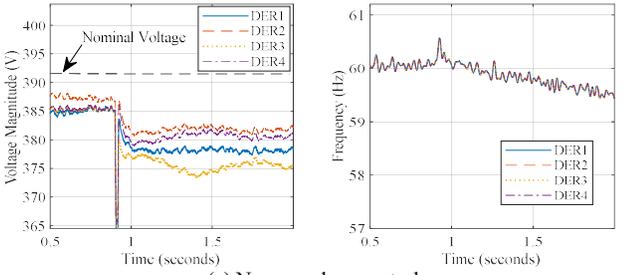
(a) No secondary control

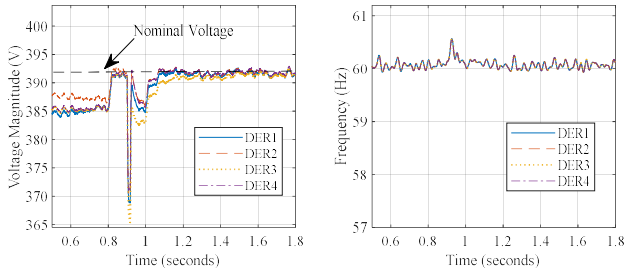
(b) The proposed AKOOC (full basis with ensemble)

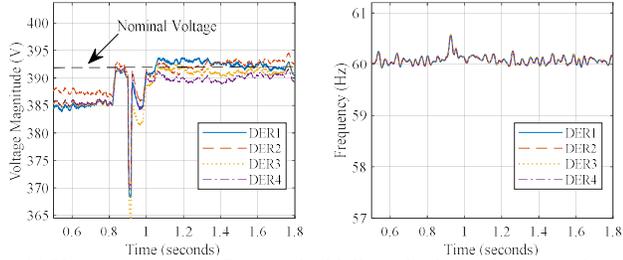
(c) Koopman-based LQR control with linear basis (equivalent to the special case of AKOOC with $C_M$ forced to $\mathbf{0}$)

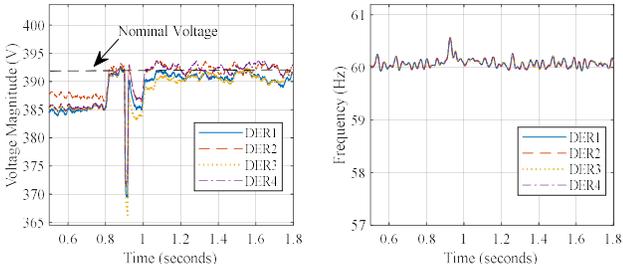
(d) Koopman-based LQR control using the full basis without ensemble

Fig. 7. Voltage and frequency trajectories of control with different basis functions with/without ensemble framework. The secondary control was enabled at 0.8s. The load at bus 3 increased at 0.9s: $P_{L2}$ increased from 19kW to 79kW and $Q_{L2}$ increased from 9kVar to 39kVar.

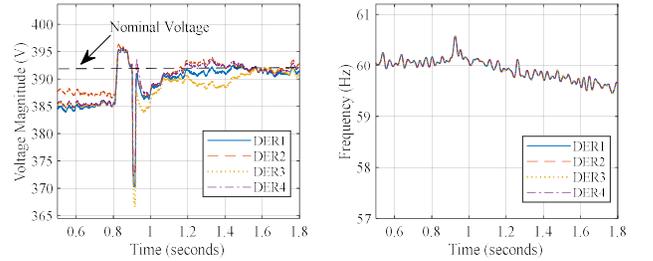
(a) Conventional secondary PI control without adaptively updated reference power

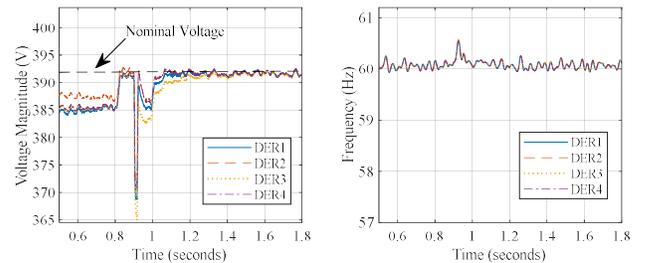
(b) Secondary PI control with adaptively updated reference power

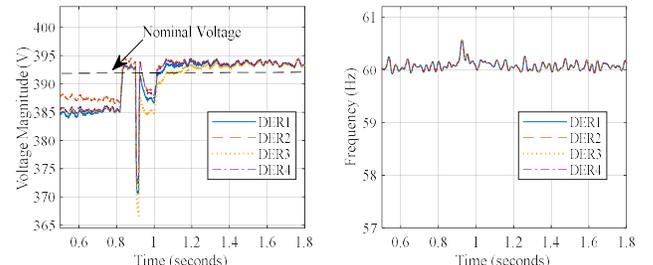
(c) Secondary droop control [17][18]

Fig. 8. Voltage and frequency trajectories with other control benchmark methods.

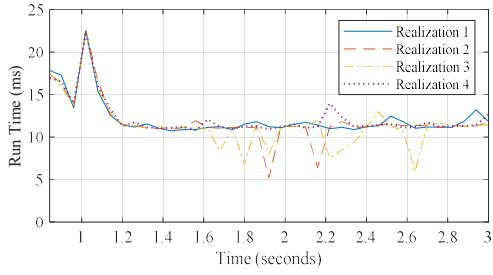

Fig. 9. Run time of the proposed AKOOC (four independent simulation realizations).

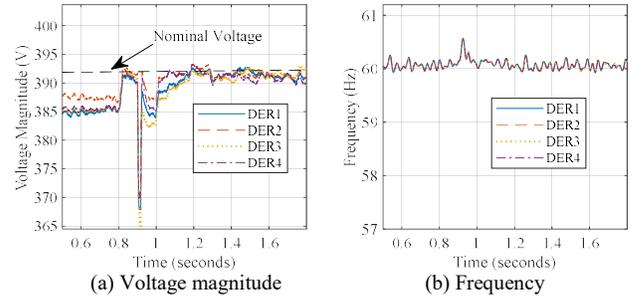

(a) Voltage magnitude  (b) Frequency

Fig. 10. The proposed AKOOC with missing PMU data at DER3.

the large disturbance at 0.9s if without any secondary control. Fig. 7(b)-(d) shows the voltage and frequency trajectories with different secondary control methods, i.e., AKOOC, Koopman-based LQR control with linear basis, Koopman-based LQR control using the full basis without ensemble. It is observed that the proposed AKOOC realizes smoother and more consistent voltage and frequency restoration to the nominal values (392V and 60Hz) than the other methods under measurement noise, control time delay, ambient perturbation, and large disturbances.

Fig. 8 shows the voltage and frequency trajectories with three other benchmark secondary control methods: (a) conventional PI control without adaptively updated reference power; (b) PI control with adaptively updated reference power; (c) secondary droop control adapted from [17], [18] based on the ingredients for voltage and frequency restoration. The terminology "with/without adaptively updated reference power" represents the reference power baseline whether or not adaptively updated at each time step according to MG operating points. That is, at each time step $k$ of secondary control, the reference power of local droop control of DERs is regulated by the secondary controller as:

$$P^* = P^{(baseline)} + \Delta P^*, Q^* = Q^{(baseline)} + \Delta Q^* \quad (37)$$

For PI control without adaptively updated reference power, $P^{(baseline)} = P_0$ and $Q^{(baseline)} = Q_0$ where $P_0$ and $Q_0$ represent the initial condition of operating points before the secondary control is enabled. For PI control with adaptively updated reference power, $P^{(baseline)} = P$ and $Q^{(baseline)} = Q$ where $P$ and $Q$ are the present low-pass-filtered power updated on a rolling basis. $\Delta P^*$ and $\Delta Q^*$ are control inputs regulated by PI control. Note that both PI control methods are tuned with the best effort from trial and error. By comparing the results in Fig. 8(a)-(b) with Fig. 7(b), we found that the PI controller without updated reference power baseline cannot achieve comparable performance for both voltage and frequency, while the PI controller with updated reference power baseline can achieve similar results to that of AKOOC. The reason is that the adaptive-updated reference power can help compensate for the nonlinearity in power flow, and the best-tuned PI control is robust to bound the rest of the modeling uncertainty. Nevertheless, advanced PI parameter tuning methods such as Ziegler-Nichols tuning and intelligent searching-based tuning require either the system's physical model or simulators to mimic the system response to conduct the tuning. In contrast, the proposed AKOOC can estimate the unknown system online without knowing the parameters and without warm-up training. In addition, AKOOC can outperform the best-tuned PI control with adaptively updated reference power in a larger MG, which will be presented in Section C. Fig. 8(c) shows that secondary droop control adapted from [17], [18] cannot restore the voltage to nominal values without the assumption of quasi-steady states between primary and secondary control.

**Run time of AKOOC.** The online estimation in AKOOC needs to be fast enough to adaptively update the ensemble Koopman model at each time step on a rolling basis. To evaluate the feasibility of such online adaptiveness of AKOOC, four independent simulation realizations were conducted on the desktop equipped with Intel Core™ i7-10700H CPU@2.90Ghz and 16G RAM, and the run time was recorded in Fig. 9. It was shown that the run time was around 12ms, which is sufficiently short compared to the secondary control time step of 60ms. This indicates the feasibility of adaptive model update at each time step with the latest measurement data.

### B. Missing PMU

Data-driven control could be affected by missing data. The AKOOC method was also tested when the PMU data at DER3 was missing. It is known that the missing state can be observed with the information at the adjacent buses [33]. Considering the model-free stochastic nature of the system and for simplicity, we approximate it with the randomly weighted averaging of the measurements of the proximal buses. For example, let $\hat{x}_{k,\text{DER3}} = (1 - rnd)x_{k,\text{DER2}} + rnd \cdot x_{k,\text{DER4}}$, where $rnd$ is a uniformly distributed random number between 0 and 1 updated each time step. As such, $0 \leq \|\hat{x}_{k,\text{DER3}}\| \leq \max\{\|x_{k,\text{DER2}}\|, \|x_{k,\text{DER4}}\|\}$, i.e., the norm of the approximation of the missing states is bounded by the maximum of the norm of the states at the adjacent DERs. From the perspective of control, the introduced stochasticity can help visit the most controllable directions and thus facilitate the reachability towards desired points in state space. Fig. 10 shows that AKOOC can restore voltage and frequency to their nominal values even with a missing PMU, indicating the robustness of the method under missing PMUs. However, the

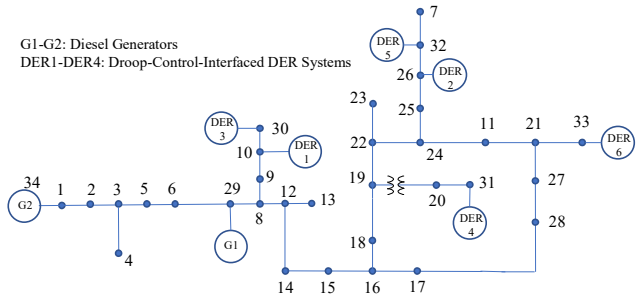

Fig. 11. Microgrid test system II (34 buses).

TABLE III.
Test System II and Control Parameters

| Parameters | | Values | Parameters | Values |
|---|---|---|---|---|
| Power base | $S_{base}$ | 150kVA | $q_\theta$ | 0 |
| Voltage base | $V_{base}$ | 12.47kV | $q_V$ | $10^2 I$ |
| Secondary control time step | $T_s$ | 60 ms | $q_{cos}$ | 0 |
| Droop gains | $\sigma_\theta$ | 2 | L  $q_{sin}$ | 0 |
| | $\sigma_V$ | 2 | Q  $q_\omega$ | $I$ |
| Droop damping parameters | $\tau_\theta$ | 8s | R  $r_P$ | $I$ |
| | $\tau_V$ | 4s | $r_Q$ | $I$ |
| Regression window size | $N$ | 10 | $U_{LB}$ | $-\frac{1}{20}$ |
| Constant for learning rate | $\gamma$ | 5 | $U_{UB}$ | $\frac{1}{20}$ |
| Maximum iteration number $N_{ITER}$ | | 250 | | |
| Regression error tolerance $\varepsilon$ | | $10^{-20}$ | | |
| Ambient active and reactive power perturbations: Wiener process | | | | |
| Dynamic load time constant: 3s | | | | |

Note: $\mathcal{N}(a, b)$ is the normal distribution with mean of $a$ and variance of $b$.

voltage trajectories are not as smooth as those without missing PMUs displayed in Fig. 7(b). If many PMUs for sensitive states are missing, the control performance can be affected.

### C. A Large Distribution Network with Dynamic Load

To test the proposed AKOOC method in a larger system, a 34-bus MG test system adapted from the IEEE 34-node test feeder was implemented in PSAT 2.1.10. Fig. 11 shows the network structure, and Table III presents the system information and control parameters. There are six droop-control-interfaced DER systems at Buses 10, 26, 30, 31, 32, and 33 (with the modified droop control in (4)), and two diesel generators located at Buses 29 and 34. The other buses are load buses with dynamic load attached. The dynamic load model can be found in [34], [35]; it can be used to characterize a variety of loads (such as thermostatic loads and induction motors). Two case studies are conducted to validate the effectiveness of the proposed AKOOC. The following case was conducted: a load perturbation occurred at 5.7s: both the active and reactive power increased by 30% at Buses 2-16, 21-23 and 25. The control started and kept online from 5.9s, i.e., 0.2s after the first load perturbation, representing a time delay in practical applications. Then, a fault occurred at Bus 26 at 16.2s, which was cleared at 16.3s with the line 25-26 disconnected.

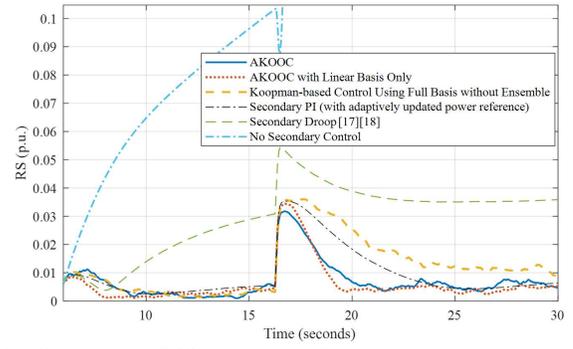

Fig. 13. Comparison of RS values for different secondary control methods. The control started at 5.9s. 30% reactive and active power increase occurred at 5.7s at Buses 2-16, 21-23, and 25. Then, a fault happened at 16.2s at Bus 26, and was cleared at 16.3s with the line 25-26 disconnected.

Fig. 12 shows the voltage deviation at the DER buses without and with the secondary control. The system may collapse without any secondary control, while the secondary control methods (i.e., AKOOC, secondary PI control with adaptively updated reference power, and secondary droop control adapted from [17], [18]) can stabilize the system while performing differently. To quantify control performance, the normalized root mean square value of the voltage and frequency deviation is defined as the index:

$$RS = \sqrt{\frac{\|\Delta V_{DER}\|_2^2 + \|\Delta \omega_{DER}\|_2^2}{2N_{DER}}} \quad (38)$$

where $N_{DER}$ is the number of DERs. A small RS value means a small state deviation. Fig. 13 shows the comparison of dynamic RS values among the following data-driven control methods: (i) the proposed AKOOC (Koopman-based LQR control using the full basis in an ensemble framework); (ii) Koopman-based LQR control with the linear basis (i.e., AKOOC with $C_M$ forced to 0); (iii) Koopman-based LQR control using the full basis $\chi$ without ensemble; (iv) secondary PI control with adaptively updated reference power; (v) secondary droop control [17], [18]. The proposed AKOOC performed similarly well with the Koopman-based LQR control using the linear basis only, outperformed the Koopman-based LQR control using the full basis without ensemble, and led to the smallest RS overshoot subject to the transient at 16.2s. It was also shown that the secondary PI and secondary droop control suffered from the sluggish responses. Besides, the secondary droop control suffered from a large steady-state control error due to the non-neglectable line impedance, the coupling of voltage and frequency, and the interaction of primary and secondary control. In short, the results indicate the generality and effectiveness of the proposed AKOOC for different MGs. The AKOOC can be deployed online to prepare for both small and large disturbances, both of which can happen frequently in modern fast-evolving distribution networks.

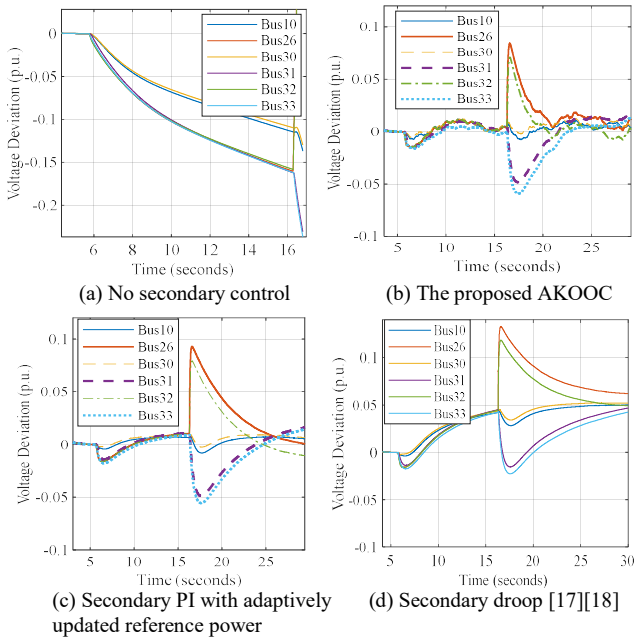

Fig. 12. Voltage deviation with/without control. The control started at 5.9s. 30% reactive and active power increase occurred at 5.7s at Buses 2-16, 21-23, and 25. Then, a fault happened at 16.2s at Bus 26, and was cleared at 16.3s with the line 25-26 disconnected.

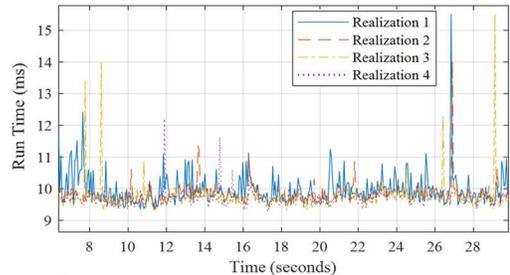

Fig. 14. Run time of the proposed AKOOC (four independent simulation realizations).

Fig. 14 shows the run time of the proposed AKOOC in four simulation realizations. The run time was around 10-15ms, shorter than the secondary control time step of 60ms. This indicates the feasibility of online applications of AKOOC.

## V. CONCLUSIONS

This paper proposed an AKOOC (Adaptive Koopman Operator Optimal Control) for MG secondary voltage and frequency control. The proposed data-driven AKOOC has an ensemble structure that can handle nonlinearity and uncertainty in MG's normal operation without offline training. A fast iterative learning method is designed to estimate the Koopman state space model adaptively online, ensuring the effectiveness of the proposed AKOOC under diverse operating conditions. With the proposed AKOOC, the system is guaranteed to be BIBO stable. On top of the BIBO stability, the sufficient condition under which the MG system is asymptotically stable is also developed. Case studies verified the effectiveness, adaptiveness, and robustness of the proposed AKOOC under varying operating conditions and topologies even with time delay, measurement noise, and missing measurements.

## APPENDIX

### A. Derivation of Ensemble Koopman State Space Model

This section derives the ensemble Koopman state space model (18) from $\hat{x}_{k+1} = (1-\beta)\hat{x}_{k+1}^{(\ell)} + \beta \hat{x}_{k+1}^{(n)}$, which is the equation (17) in Section III.C.

Suppose that $B_M = \widehat{B}_M + H_M$; pre-multiply $C_M$ and we have:

$$B = C_M B_M = C_M \widehat{B}_M + C_M H_M = C_M \widehat{B}_M + H \quad (39)$$

where $H_M$ and $H$ are error matrices for the estimation satisfying $H = C_M H_M$. They are bounded as the mapping approximation in (13) has the bounded errors $e^{(L)}$ and $e^{(NL)}$. The uncertainty can be compensated by an online data-driven Koopman-based ensemble framework. Suppose that a linear controller is applied to the Koopman embeddings: $u_k = K \chi_k$ with a constant control gain matrix $K = [K_X, K_Z]$; i.e., $u_k = K_X x_k + K_Z z_k$. Then according to (14)-(17) and (39), we know

$$\hat{x}_{k+1} = (1-\beta)(Ax_k + Bu_k) + \beta C_M(A_M z_k + \widehat{B}_M u_k + H_M u_k)$$
$$= (1-\beta)Ax_k - \beta B u_k + B u_k + C_M(\beta A_M z_k - (1-\beta)\widehat{B}_M u_k + \widehat{B}_M u_k + \beta H_M u_k) = ((1-\beta)A - (B-\beta H)K_X)x_k + B u_k + C_M((\beta A_M - (\widehat{B}_M - \beta H_M)K_Z)z_k + \widehat{B}_M u_k) \quad (40)$$

For convenience, denote the linear-based and nonlinear-based models in Fig. 2 by $M_1(x_k, u_k; A_E) \coloneqq A_E x_k + B u_k$ and $M_2(z_k, u_k; A_{EM}) \coloneqq A_{EM} z_k + \widehat{B}_M u_k$, respectively. If define $A_E \coloneqq (1-\beta)A - (B - \beta H)K_X$ and $A_{EM} \coloneqq \beta A_M - (\widehat{B}_M - \beta H_M)K_Z$, we can rewrite (40) as:

$$\hat{x}_{k+1} = M_E(x_k, z_k, u_k; W) = M_1(x_k, u_k; A_E) + C_M M_2(z_k, u_k; A_{EM}) \quad (41)$$

where $W = \{A_E, A_{EM}\}$ is the set of new matrix parameters to be learned, having the unknown state transition matrices $A$ and $A_M$ in (14) and (15), the unknown ensemble weight $\beta$ and the uncertainty matrices $H$ and $H_M$ incorporated in a compact form. Instead of estimating each of these parameters, we focus on estimating the two matrix parameters $W = \{A_E, A_{EM}\}$, which can be learned online with small data.

**The Koopman State Space Model in the Ensemble Framework.** Let $\delta_k^{(x)}$ be the error term of $x$ with the new ensemble model (41), we have

$$x_{k+1} = \hat{x}_{k+1} + \delta_k^{(x)} = A_E x_k + B u_k + C_M(A_{EM} z_k + \widehat{B}_M u_k) + \delta_k^{(x)} \quad (42)$$

Conducting the mapping from $x$ to $z$ for (42) by pre-multiplying $C_M^\dagger$ on both sides and according to (13), we have

$$\hat{z}_{k+1} = C_M^\dagger A_E x_k + C_M^\dagger B u_k + (A_{EM} z_k + \widehat{B}_M u_k) + C_M^\dagger \delta_k^{(x)}$$
$$\Rightarrow z_{k+1} = C_M^\dagger A_E x_k + A_{EM} z_k + (\widehat{B}_M + C_M^\dagger B) u_k + \delta_k^{(z)} \quad (43)$$

where $\delta_k^{(z)} = C_M^\dagger \delta_k^{(x)} + e^{(NL)}$ is the modeling error term of $z$. Combining (42)-(43), we obtain the ensemble Koopman state space model (18) with the Koopman modeling error $\delta_{E,k} = \left[\left(\delta_k^{(x)}\right)^T, \left(\delta_k^{(z)}\right)^T\right]^T$.

### B. Effectiveness and Error Bounds of Ensemble Koopman State Space Learning

Given dataset $\{x, u\} \in D_k$, let $Y_S(A_E)$ and $Y_L(A_{EM})$ be the matrices for the individual models $M_1$ and $M_2$:

$$Y_S(A_E) = \begin{bmatrix} | & & | \\ M_1(x_{k-N+1}, u_{k-N+1}; A_E), & \dots & M_1(x_k, u_k; A_E) \\ | & & | \end{bmatrix} \quad (44)$$

$$Y_L(A_{EM}) = \begin{bmatrix} | & & | \\ M_2(z_{k-N+1}, u_{k-N+1}; A_{EM}), & \dots & M_2(z_k, u_k; A_{EM}) \\ | & & | \end{bmatrix} \quad (45)$$

Then we define $\delta_{M1} \coloneqq X' - Y_S(A_E)$ and $\delta_{M2} \coloneqq X' - C_M Y_L(A_{EM})$, where $\delta_{M1}$ and $\delta_{M2}$ are the matrices of modeling residuals for the individual models $M_1$ and $M_2$, respectively. Denoted by $\delta^{(X)}$ the matrix of the ensemble modeling error defined in (42). Then

$$\delta^{(X)} = X' - Y_S(A_E) - C_M Y_L(A_{EM})$$
$$= X' - (X' - \delta_{M1}) - C_M Y_L(A_{EM}) = \delta_{M1} - C_M Y_L(A_{EM}) \quad (46)$$
$$\delta^{(X)} = X' - Y_S(A_E) - C_M Y_L(A_{EM})$$
$$= X' - Y_S(A_E) - (X' - \delta_{M2}) = \delta_{M2} - Y_S(A_E) \quad (47)$$

The least-squares optimization in (23) is to minimize the Frobenius norm of $\delta^{(X)}$. That is

$$\min \|\delta^{(X)}\|_F = \min_{A_E, A_{EM}} \left( \|\alpha(\delta_{M2} - Y_S(A_E)) + (1-\alpha)(\delta_{M1} - C_M Y_L(A_{EM}))\|_F \right)$$
$$\leq \alpha \left( \min_{A_E} \|\delta_{M2} - Y_S(A_E)\|_F \right) + (1-\alpha)\left( \min_{A_{EM}} \|\delta_{M1} - C_M Y_L(A_{EM})\|_F \right), \text{ for all } \alpha \quad (48)$$

Then iterative learning is conducted between the above two minimization terms (i.e., $\min_{A_E}\|\delta_{M2} - Y_S(A_E)\|_F = \min_{A_E}\|X' - Y_S(A_E) - C_M Y_L\|_F$ and $\min_{A_{EM}}\|\delta_{M1} - C_M Y_L(A_{EM})\|_F = \min_{A_{EM}}\|X' - Y_S - C_M Y_L(A_{EM})\|_F$). This exploits a local minimum which can be seen as an equilibrium of a two-player cooperative game [36] between $M_1$ and $M_2$. Such iterative learning can also improve numerical conditioning for the augmented space $\chi = [x^T, z^T]^T$. Now, denoted by $\epsilon_1$ and $\epsilon_2$, the error bounds for the two minimization terms in (48). We have

$$\min\|\delta_k^{(x)}\|_2 \leq \min\|\delta^{(X)}\|_F \leq \alpha\epsilon_1 + (1-\alpha)\epsilon_2, \text{ with } \epsilon_1 > 0, \epsilon_2 > 0 \quad (49)$$

which means that the modeling error for $x_k = [\Delta\theta_k, \Delta V_k]^T$ are bounded by a small positive value. Suppose the $i^{th}$ Koopman embedding function $g_i(x)$ is $L$-Lipschitz continuous, i.e., there exists an arbitrary constant $L$ making $\|g_i(x + \varsigma) - g_i(x)\|_2 \leq L \cdot \|\varsigma\|_2$ hold for any given arbitrary value $\varsigma$. Then the modeling error $\delta_{E,k}$ for Koopman embeddings $\chi_k = g(x_k)$ are also bounded, which is

$$\|\delta_{E,k}\|_2 \leq L \cdot \|\delta_k^{(x)}\|_2 \leq \alpha\epsilon_1 L + (1-\alpha)\epsilon_2 L \quad (50)$$

In addition, over the iteration learning, the following inequalities hold:

$$\min\|\delta^{(X)}\|_F = \min_{A_M}\|\delta_{M1} - C_M Y_L(A_{EM})\|_F \leq \|\delta_{M1}\|_F \quad (51)$$

$$\min\|\delta^{(X)}\|_F = \min_A \|\delta_{M2} - Y_S(A_E)\|_F \leq \|\delta_{M2}\|_F \quad (52)$$

that the ensemble model can yield a smaller error compared to each individual model $M_1$ or $M_2$.

### C. Stability of AKOOC

The AKOOC is BIBO (bounded-input-bounded-output) stable.

*Proof.* Denoted by $x_{k+1}^-$ the one step-ahead prediction of the state vector $x$ at the time step $k$. Denoted by $\widetilde{\mathcal{K}}_k$ the estimated Koopman operator at time step $k$. According to (10), we have

$$g(x_{k+1}^-) = \Phi(x_{k+1}, 0) = \widetilde{\mathcal{K}}_k \Phi(x_k, u_k)$$
$$= \widetilde{\mathcal{K}}_k \sum_{j=1}^{N_\psi} \psi_j(x_k, u_k)v_j = \sum_{j=1}^{N_\psi}(\lambda_{j,k}\psi_j(x_k, u_k)v_j) \quad (53)$$

where $\lambda_{j,k}$ is the eigenvalue corresponding to the $j^{th}$ eigenfunction $\psi_j$ for the estimated Koopman operator $\widetilde{\mathcal{K}}_k$. According to Section III.A, recall that $\Phi(x, u) = g(x) + L(u)$ where $L(u) = [L_1(u), L_2(u), \ldots L_p(u)]^T$ and $L(0) = 0$. Then

$$g(x_{k+1}) = g(x_{k+1}^-) + \delta_{E,k} = \widetilde{\mathcal{K}}_k \Phi(x_k, u_k) + \delta_{E,k}$$
$$= \widetilde{\mathcal{K}}_k(\Phi(x_k, 0) + L(u_k) + \delta_{E,k-1}) + \delta_{E,k}$$
$$= \widetilde{\mathcal{K}}_k(\widetilde{\mathcal{K}}_{k-1}\Phi(x_{k-1}, u_{k-1}) + L(u_k) + \delta_{E,k-1}) + \delta_{E,k}$$
$$= \prod_{h=0}^k \widetilde{\mathcal{K}}_{k-h}\Phi(x_0, u_0) + \sum_{h=1}^k \prod_{i=h}^k \widetilde{\mathcal{K}}_{k-i+h}(\delta_{E,h-1} + L(u_h)) + \delta_{E,k}$$
$$= \sum_{j=0}^{N_\psi}(\prod_{h=0}^k \lambda_{j,h})\psi_j(x_0, u_0)v_j + \sum_{h=1}^k \prod_{i=h}^k \widetilde{\mathcal{K}}_{k-i+h}(\delta_{E,h-1} + L(u_h)) + \delta_{E,k} \quad (54)$$

where $\delta_{E,k}$ is the ensemble Koopman modeling error, and $v_j$ is the $j^{th}$ Koopman mode associated with the Koopman eigenfunction $\psi_j$. Apparently,

$$0 \leq \left\|\sum_{j=0}^{N_\psi}(\prod_{h=0}^k \lambda_{j,h})\psi_j(x_0, u_0)v_j\right\|_2 \leq \lim_{k\to\infty}\left(\max_{j,h}|\lambda_{j,h}|\right)^{k+1}\sum_{j=1}^{N_\psi}\|\psi_j(x_0, u_0)v_j\|_2 \quad (55)$$

With LQR in the Koopman invariant subspace, the MG secondary controller can optimally make the magnitudes of all system eigenvalues smaller than 1 (if the system is stabilizable) [37]. That is, $\widetilde{\mathcal{K}}_k\psi_j = \lambda_{j,k}\psi_j$ with $|\lambda_{j,k}| < 1$. In AKOOC, due to the rolling-based estimation, we assume the global error $\left\|\sum_{h=1}^k \prod_{i=h}^k \widetilde{\mathcal{K}}_{k-i+h}(\delta_{E,h-1} + L(u_h))\right\|_2$ is bounded by $\rho_g$. According to (54), we have

$$0 \leq \lim_{k\to\infty}\|g(x_{k+1})\|_2 \leq \lim_{k\to\infty}\left(\sum_{j=0}^{N_\psi}\left(\max_{j,h}|\lambda_{j,h}|\right)^{k+1}\sum_{j=1}^{N_\psi}\|\psi_j(x_0, u_0)v_j\|_2\right)$$
$$+\rho_g + \lim_{k\to\infty}\|\delta_{E,k}\|_2 \leq 0 + \rho_g + \alpha\epsilon_1 L + (1-\alpha)\epsilon_2 L, \text{ for all } \alpha$$
$$\Rightarrow 0 \leq \lim_{k\to\infty}\|g(x_{k+1})\|_2 \leq \rho_g + \epsilon_2 L + \min_\alpha(\alpha L(\epsilon_1 - \epsilon_2)) \quad (56)$$

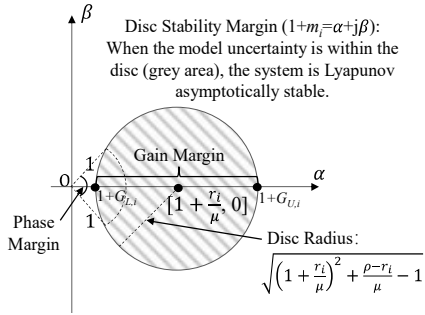

Fig. 15. Disc stability margin for the discrete-time LQR.

Based on (56), $g(x)$ converges till reaching the area $\Theta = \{g(x) \mid \|g(x)\|_2 \leq \rho_g + \epsilon_2 L + \min_\alpha(\alpha L(\epsilon_1 - \epsilon_2))\}$. Thus, the system is BIBO stable. Besides, the Koopman-enabled LQR can guarantee asymptotic stability subject to the disturbance in control input channels under mild conditions. See Appendix D.

### D. Stability Margins of Koopman-enabled LQR

The discrete-time LQR used in this paper has analytical disc stability margins [38], within which asymptotic stability subject to the disturbance in control input channels is guaranteed. Specifically, consider the identified ensemble Koopman state space model described as below:

$$g(x_{k+1}) = A_C g(x_k) + B_C u_k + B_C M u_k$$
$$= A_C g(x_k) + B_C K g(x_k) + B_C M K g(x_k)$$
$$= A_C g(x_k) + B_C (I + M) K g(x_k) \quad (57)$$

where $M = \text{diag}([m_1, m_2, \ldots m_{2N_{DER}}])$ is an introduced diagonal matrix to represent model uncertainty in control input channels, and $K$ is the control gain matrix. We consider $M$ and $g(x_k)$ to be complex-valued to reflect both gain and phase disturbances. Following the steps in [38] with a Lyapunov function defined as $V(x) = g(x)^* S g(x)$ (where $S$ is the solution of Riccati equation), one can get the disk stability margin for the $i^{th}$ control input channel as follows:

$$1 + m_i = \left\{\alpha + j\beta : \left(\alpha - \left(1 + \frac{r_i}{\mu}\right)\right)^2 + \beta^2 < \left(1 + \frac{r_i}{\mu}\right)^2 + \frac{\rho - r_i}{\mu} - 1\right\}$$
$$\text{where } i = 1, 2, \ldots, 2N_{DER} \quad (58)$$

where $\rho = \sigma_{min}[Q]/(\sigma_{max}[K])^2$ and $\mu = \sigma_{max}[B_C^T S B_C]$. $\sigma_{max}[\cdot]$ and $\sigma_{min}[\cdot]$ represent the matrix operation to obtain the maximum and minimum singular values, respectively. $r_i$ is the $i^{th}$ diagonal element of the cost matrix $\mathcal{R}$. Fig. 15 shows the disc margin. Based on the disc margin, we derive the *sufficient conditions* of asymptotic convergence against the model uncertainty in what follows.

***Sufficient condition of asymptotic stability subject to the disturbance in control input channels.*** The ensemble Koopman state space model (57) is updated at each time step in AKOOC; therefore, the gain margin is of interest. Apparently according to Fig. 15, if the diagonal elements of $I + M$ are within the gain margin, i.e., $1 + G_{L,i} < 1 + m_i < 1 + G_{U,i}$ with $G_{L,i}$ and $G_{U,i}$ as follows

$$G_{L,i} = \frac{r_i}{\mu} - \sqrt{\left(1 + \frac{r_i}{\mu}\right)^2 + \frac{\rho - r_i}{\mu} - 1}, \quad G_{U,i} = \frac{r_i}{\mu} + \sqrt{\left(1 + \frac{r_i}{\mu}\right)^2 + \frac{\rho - r_i}{\mu} - 1}$$
$$\text{for } i = 1, 2, \ldots, 2N_{DER} \quad (59)$$

then the model is asymptotically stable subject to the disturbance in control input channels.

That is, the *sufficient condition* of asymptotic stability subject to the disturbance in control input channels is that the diagonal elements of uncertainty matrix $M$ satisfy:

$$G_{L,i} < m_i < G_{U,i}, \quad \text{for } i = 1, 2, \ldots, 2N_{DER} \quad (60)$$

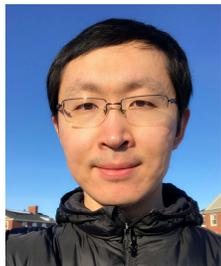
**Xun Gong** (S'15-M'21) is currently a Postdoctoral Researcher in the Department of Electrical and Computer Engineering at McGill University, Montreal, QC, Canada. He received the Ph.D. degree from the Emera and NB Power Research Center for Smart Grid Technologies, University of New Brunswick, Fredericton, Canada, in 2021, and received B.S.E.E and M.Sc. from Hefei University of Technology, China, in 2011 and 2014, respectively. His research interests include data-driven control and optimization for power systems, energy modeling and forecasting, demand response, and microgrids.

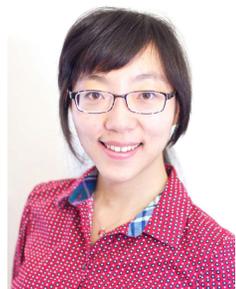
**Xiaozhe Wang** (S'13-M'15-SM'20) is currently an Associate Professor in the Department of Electrical and Computer Engineering at McGill University, Montreal, QC, Canada. She received the Ph.D. degree in the School of Electrical and Computer Engineering from Cornell University, Ithaca, NY, USA, in 2015, and the B.S. degree in Information Science & Electronic Engineering from Zhejiang University, Zhejiang, China, in 2010. Her research interests are in the general areas of power system stability and control, uncertainty quantification in power system security and stability, and wide-area measurement system (WAMS)-based detection, estimation, and control. She is serving on the editorial boards of IEEE Transactions on Power Systems, Power Engineering Letters, and IET Generation, Transmission and Distribution.

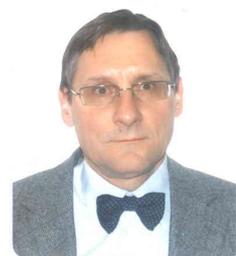
**Geza Joos** (F'2004) graduated from McGill University, Montreal, Canada, with an M.Eng. and Ph.D. in Electrical Engineering. He is a Professor in the Department of Electrical and Computer Engineering Department at McGill University (since 2001). He holds a Canada Research Chair in Powering Information Technologies (since 2004). His research interests are in distributed energy resources, including renewable energy resources, advanced distribution systems and microgrids. He was previously with ABB, the Université du Québec and Concordia University (Montreal, Canada). He is active in IEEE Standards Association working groups on distributed energy resources and microgrids. He is a Fellow of CIGRE, and the Canadian Academy of Engineering.